# THE 'POWER OF *THEN*':

## The Uniquely Human Capacity to Imagine Beyond the Present

Liane Gabora

Department of Psychology, University of British Columbia
Okanagan Campus, Arts Building, 333 University Way, Kelowna BC, V1V 1V7, CANADA

**Abstract**
Humans appear to be uniquely in their ability to transcend the present and reflect on current ideas in terms of what was experienced in the past, or fantasize about the future. This paper presents an account, in layperson terms, of how this ability came about, its importance in modern life, and why it defines our 'human-ness'.

*Live in the present*. This is widely accepted as ancient wisdom, as the key to escaping the speed and pressures of modern life. The notion of 'living in the present' has become even more popular with the publication of *The Power of Now,* a New York Times number one bestseller by Eckhart Tolle that has been translated into 30 languages. We are told that authentic human power comes by surrendering to the Now, the present moment, which is laden with myriad access points or portals to a state of inner peace where problems do not exist. We are told that when we dwell in the Now, we leave behind our analytic mind and its falsely created self, the ego. We may thereby awaken spiritual dimensions of our lives, enjoy more harmonious relationships, and discover that we are already complete and perfect.

It may be that we can use a reminder now and then to forget our worries, appreciate our surroundings, and simply 'be'. In so doing, inner strength may be found. But it is equally true that we can leave behind the ego and awaken new dimensions in our lives by *escaping* the now, by dwelling in the 'Then'. There is good reason why, after millions of years of evolution, humans became able to escape the present, i.e. to plan, imagine, fantasize, and remember past events without a memory-jogging cue. Indeed when our ancestors acquired the capacity to break out of the present, it opened up for our species an entirely unprecedented and breathtakingly creative new kind of existence.

**Abstract Thought: The Capacity to Transcend the Present**

Other than rocks and plants and things of the natural world, everything around you began as a spark of insight that came to its creator's mind in a moment of *transcending the Now*. As far as we know, no other species is able to envision something that does not exist. Cognitive scientist Merlin Donald (1991) refers to the minds of other species--and the minds our own early ancestors--as *episodic*. The awareness of an episodic mind, he claims, is confined to the 'here and now' that is delivered by the senses. Occasionally something in an animal's surroundings reminds it of something from the past, or stimulates a desire for something in the immediate future. For example, if a deer encounters a grizzly bear in the wild, this might rekindle the memory of a previous grizzly encounter, and the experience might linger for a while as it gets cemented in long term memory. But we have no evidence that the deer will thereafter mull over different potential techniques for escaping a bear, perhaps comparing and contrasting this with methods for escaping other predators. It will certainly never recount the experience to other deer, or write a story or a song inspired by the grizzly encounter. Similarly it might see a potential mate and have sexual cravings, but not build an elaborate fantasy world around the object of its desire.

Humans, in contrast, are not only able to transcend the present, but compelled to do so. Our minds shift smoothly between past, present and future almost without our being aware of it. The shower or drive to work is seamlessly woven with flights of fancy: excursions into what might be going on in someone else's mind, or perhaps what it would take to make a certain product appeal to a new segment of the market. Or, as you wash your hair, you may be, silently rehearsing a piece of music, or lecture, or basketball move. We escape the present by engaging in *abstract thought*. Abstract thought un-tethers us from the realm of 'what is', the here and now conveyed by our senses, and transports us to the realm of 'what could be', the playing field of our imagination. It enables us to not only envision things that do not exist, but actualize things we envision, bring them into existence. It underlies works of creative genius and acts of global devastation alike. How did the ability to engage in abstract thought come about?

The split between humans and the great apes occurred some five to ten million years ago. Around two million years ago *Homo erectus* appeared on the scene. Unlike their predecessors, *Homo habilis,* who generally lumbered about on all four limbs, *Homo erectus* walked upright, and had a significantly larger brain. Their appearance on this planet is widely believed to mark the beginnings of human culture: the onset of deliberate use of fire, manufacture of tools, strategic hunting in organized groups, and migration out of Africa. To leave the homeland would have required our ancestors to adapt their food preferences, hunting techniques, methods of building shelter, averting natural disaster, and so forth, to match the possibilities and dangers offered by the new environment. What made our ancestors suddenly capable of all this?

Merlin Donald (1991) claims that it was the appearance of what he refers to as a 'self-triggered recall and rehearsal loop'. That is, the human brain became able to recursively activate items from memory, such that one triggers another, which triggers another, and so forth. Although the initial thought might be of something in the Now--perhaps a need or desire, or something in the present surroundings--each thought triggers another that offers a glimpse of something increasingly far removed from the present. This ability to string thoughts together to form a 'stream of thought' enabled us to relive events that happened previously and bring them to life for each other, initially just through mime, but eventually through the spoken word as well. It enabled us to rehearse and refine actions and ideas, modifying them progressively, one step at a time. With this 'self-triggered recall and rehearsal loop' we could now activate and re-activate visions and dreams, such that with each successive conception of them they were looked at from a different angle, embedded a little more firmly in the constraints of reality as we know it, and potentially turned into a form in which they could be realized. It is this that has transformed the world into a place where we are surrounded by objects that are as much the product of human imagination as of nature. We have become immune to the wonder of this, but to our early ancestors, the ability to create something in the mind, and then make it real--turn it into something tangible in the world--must have seemed miraculous.

Thus the onset of this 'self-triggered recall and rehearsal loop' made it possible for us to engage in abstract thought, enabling us to invent, adapt, and refine ideas--in short, escape the limitations of the present. But this leads to the question: what made self-triggered recall and rehearsal possible? Surely the Homo erectus' larger brain must have had something to do with it, but why would a bigger brain allow for abstract thought? Indeed the onset of the capacity for abstract thought presents a chicken-and-egg paradox. Until items in memory are woven into an integrated internal model of the world, or *worldview,* how does one evoke another, which evokes another, and so forth, in a stream of thought? Conversely, until a mind can generate a stream of thought, how does it integrate items into a worldview? How could something composed of complex, mutually interdependent parts come to be?

**Weaving an Integrated Internal Model of the World**
This is one of the questions that has inspired my own work. It turns out that the chicken-and-egg paradox can be resolved using the notion of self-organization, which can be explained with a simple analogy that was first used by biologist Stuart Kauffman (Kauffman, 1993). Imagine that you spill a jar full of buttons on the floor. You randomly choose any two buttons and tie two them together with a thread. You repeat this again and again, picking up two buttons at random and tying them together. Occasionally you lift a button and see how many connected buttons get lifted. What you find is that clusters of buttons start to emerge. At a certain point you pass a threshold referred to as the *percolation threshold,* whereupon a giant cluster of buttons suddenly emerges. In this giant cluster button is connected, not directly but indirectly, to any other button.

Applying this to the mind, imagine that the buttons are experiences or items of information encoded in memory, and the strings are associations between them, ways in which one item can remind us of another (Gabora, 1998). The giant cluster is thus an integrated internal model of the world, or worldview. It is *self-organizing* because its structure emerges through local, 'bottom-up' interactions rather than being dictated by 'top-down' instructions. It constantly revises itself, framing new situations and ideas in terms of old ones, thereby achieving a richer understanding of them (and a better chance of remembering them). A worldview is also *self-mending* in that if you encounter new information that conflicts with your current web of understandings, you almost can't help but engage in abstract thought, reflecting on the inconsistency from different angles until the dissonance is resolved (Gabora, 1999, 2013). The worldview of one person is similar to that of another in the general sense that we all possess, for example, the concepts DOG and BEAUTY. But in the details of how we weave these concepts together, each individual's worldview is unique, and thereby offers unique possibilities for unearthing gems from the realm of 'what could be' and realizing them in the realm of 'what is'.

**The Spectrum of Abstract Thought**
Psychologists from William James to Freud noticed that there are different genres of abstract thought, or more accurately, that thought varies along a spectrum. At one end is *associative thought,* which is conducive to induction, analogy, seeing things new ways, and detecting relationships of correlation. If you write a poem in which the roots of a tree become a metaphor for the unconscious, you are exploiting the power of associative thought. Associative thought was referred to by Freud as primary process thinking, and psychologists sometimes call it divergent thought. At the other end of the spectrum is *analytic thought,* which is conducive to deduction, manipulating symbols, and figuring out relationships of cause and effect. If you know that all trees have roots and deduce that a fir tree has roots, you are exploiting the power of analytic thought. Analytic thought was referred to by Freud as secondary process thinking.

Power of Now advocates claim that the ego is the product of analytic thought. However, the ego, as Freud described it, is a structure that attempts to forge a relationship between the id, which is focused on selfish needs and desires, and the constraints of external reality. Both the ego and the id can be as active when one is absorbed in the Now as when one is catapulted into the Then. Exploring the realm beyond the Now involves both associative and analytic thought; indeed creativity requires the capacity to spontaneously and subconsciously shift between them depending on the situation. For example, if you come up with the idea of building a fence out of old skis, you are engaging in associative thought. But in order to carry it out, you have to figure out how to hammer the ski poles into the ground, and what to use as crossbeams. For that you shift into a more analytic mode of thought. You might again shift back into a more associative thought when you come up with the idea of placing white rocks at the base of the fence, reminiscent of snow. We are able to shift between these two modes of thought depending on the demands of the situation. I have proposed that the onset of the capacity to do so is what lay behind the sudden proliferation of art, science and religion in the Middle-Upper Paleolithic, some fifty thousand years ago (Gabora, 2003, 2008; Gabora & Kaufman, 2010; Gabora & Russon, 2011), a time referred to by anthropologist Steven Mithen (1998) as the "big bang of human creativity".

**The Magic of Glimpsing Beyond the Now**
'Power of Now' advocates claim that much of our thinking is repetitive and pointless, that it merely recreates and solidifies for ourselves the same old nightmarish existence. There may be a grain of truth to this, but change can be subtle. It may appear at first glance that the earth is simply rotating repetitively around the sun, but from the perspective of the entire solar system, its distance to other planets is changing, and from the perspective of the universe at large, myriad patterns of interrelationship are thereby unfolding. Upon superficial observation thoughts may appear repetitive and pointless, but more in-depth observation can reveal that they are honing in on more accurate conceptions of things, or their relationships to each other, or to ourselves. We forge a deeper understanding of the past by revisiting it in light of the present, and vice versa: by infusing the present with historical antecedents, a richer understanding of it is ultimately achieved. We encounter this not just in the workings of our own minds but also in film and literature, where elements of past, present, and future are frequently intertwined to yield a more multifaceted conception of reality. Indeed although our wiring may sometimes go awry, we are wired up in such a way as to cultivate, not interfere with, our spiritual evolution. We may not always feel 'complete and perfect', but our ability to, now and then, leave the Now behind, is a profound aspect of who we are. It helps us find meaning in daily experience, and makes us interesting.

# Acknowledgments

The author is grateful for funding from the Natural Sciences and Engineering Research Council of Canada.

Gabora, L. (1998). Autocatalytic closure in a cognitive system: A tentative scenario for the origin of culture. *Psycoloquy, 9*(67).

Gabora, L. (1999). Weaving, bending, patching, mending the fabric of reality: A cognitive science perspective on worldview inconsistency. *Foundations of Science, 3*(2), 395-428.

Gabora, L. (2003). Contextual focus: A cognitive explanation for the cultural transition of the Middle/Upper Paleolithic. *Proceedings Annual Conference Cognitive Science Society* (pp. 432-437), Boston: Lawrence Erlbaum.

Gabora, L. (2008). Mind. In (R. A. Bentley, H. D. G. Maschner, & C. Chippindale, Eds.) *Handbook of Archaeological Theories* (pp. 283-296). Walnut Creek CA: Altamira Press.

Gabora, L., & Kaufman, S. (2010). Evolutionary perspectives on creativity. In (J. Kaufman & R. Sternberg, Eds.) *The Cambridge Handbook of Creativity* (pp. 279-300). Cambridge UK: Cambridge University Press.

Gabora, L., & Russon, A. (2011). The evolution of human intelligence. In (R. Sternberg & S. Kaufman, Eds.) *The Cambridge Handbook of Intelligence*, (pp. 328-350). Cambridge UK: Cambridge University Press.

Gabora, L. (2013). An evolutionary framework for culture: Selectionism versus communal exchange. *Physics of Life Reviews, 10*(2), 117-145.

Kauffman, S. (1993). *Origins of order*. Oxford University Press.

Mithen, S. (1998). *Creativity in human evolution and prehistory*. London, UK: Routledge.

Tolle, E. (1997). *The power of now*. Namaste Publishing.